\documentclass[conference]{IEEEtran}
\usepackage[right=0.64in,left=0.63in,top=0.95in,bottom=0.99in]{geometry}
\usepackage{multirow}
\usepackage{booktabs}
\usepackage{amsfonts}
\usepackage{placeins}
\usepackage{array}
\usepackage{amsmath,cases,amssymb}
\usepackage{amsmath}
\usepackage{graphicx}
\usepackage{subcaption}
\usepackage{cite}
\usepackage{epsfig}
\usepackage{todonotes}
\usepackage{url}
\usepackage{algorithm}
\usepackage{algorithmic}
\usepackage{epstopdf}
\usepackage{balance}
\usepackage{color}
\usepackage{algorithm}
\DeclareGraphicsExtensions{.eps,.pdf}

\usepackage{scalerel}
\usepackage{multicol}
\usepackage{amssymb}
\usepackage{pifont}
\usepackage{cases}

\usepackage{scalerel}

\let\labelindent\relax
\usepackage{enumitem}

\allowdisplaybreaks

\newcommand{\bseq}{\begin{subequations}}
\newcommand{\eseq}{\end{subequations}}
\newcommand{\baln}{\begin{align}}
\newcommand{\ealn}{\end{align}}
\newcommand{\balnd}{\begin{aligned}}
\newcommand{\ealnd}{\end{aligned}}
\newcommand{\beq}{\begin{equation}}
\newcommand{\eeq}{\end{equation}}
\newcommand{\beqn}{\begin{eqnarray}}
\newcommand{\eeqn}{\end{eqnarray}}
\newcommand{\beqno}{\begin{eqnarray*}}
\newcommand{\eeqno}{\end{eqnarray*}}
\newcommand{\bma}{\begin{displaymath}}
\newcommand{\ema}{\end{displaymath}}
\newcommand{\bnu}{\begin{enumerate}}
\newcommand{\enu}{\end{enumerate}}
\newcommand{\bce}{\begin{center}}
\newcommand{\ece}{\end{center}}
\newcommand{\btb}{\begin{tabular}}
\newcommand{\etb}{\end{tabular}}
\newcommand{\bieq}{\begin{IEEEeqnarray}}
\newcommand{\eieq}{\end{IEEEeqnarray}}

\newtheorem{mypro}{\bf Proposition}

\newcommand{\st}{{\mathrm{s.t.}}}

\newcommand{\non}{\nonumber}
\newcommand{\subnum}{\IEEEyessubnumber}

\usepackage{scalerel}

\usepackage{caption}

\makeatletter
\newcommand{\linebreakand}{%
\end{@IEEEauthorhalign}
\hfill\mbox{}\par
\mbox{}\hfill\begin{@IEEEauthorhalign}
}
\makeatother

\begin{document}
\title{
Seamless 5G Automotive Connectivity with Integrated Satellite Terrestrial Networks in C-Band \vspace{-2mm}}

\author{\IEEEauthorblockN{Hung Nguyen-Kha${}^{\dagger}$, Vu Nguyen Ha${}^{\dagger}$, Eva Lagunas${}^{\dagger}$, Symeon Chatzinotas${}^{\dagger}$, and Joel Grotz${}^{\ddagger}$}

\IEEEauthorblockA{\textit{${}^{\dagger}$Interdisciplinary Centre for Security, Reliability and Trust (SnT), University of Luxembourg, Luxembourg} \\
       \textit{${}^{\ddagger}$SES S.A., Luxembourg}}
       
       \vspace{-7mm}
       }

\maketitle

\begin{abstract} 
	This paper examines integrated satellite-terrestrial networks (ISTNs) in urban environments, where terrestrial networks (TNs) and non-terrestrial networks (NTNs) share the same frequency band in the C-band which is considered the promising band for both systems. The dynamic issues in ISTNs, arising from the movement of low Earth orbit satellites (LEOSats) and the mobility of users (UEs), are addressed. The goal is to maximize the sum rate by optimizing link selection for UEs over time. To tackle this challenge, an efficient iterative algorithm is developed. Simulations using a realistic 3D map provide valuable insights into the impact of urban environments on ISTNs and also demonstrates the effectiveness of the proposed algorithm.
\end{abstract}
\vspace{-1mm}



\section{Introduction}
\vspace{-1mm}
Recently, ISTNs have emerged as a promising solution for achieving seamless connectivity and meeting high data rate demands \cite{SurveyTut_Roadto6G, ProIEEE_6GVision_Challenge_Opp,Zaid_PIMRC23,fontanesi2023artificial,Hung_WSA23, hung_ICCW_2023twotier, Hung_2tierLEO}. 
Particularly, as the connection number and the service demand increase rapidly, especially in urban areas, TNs struggle to ensure a high data rate for massive connectivity. Additionally, ensuring TN coverage in urban areas requires dense BS deployment which is a difficult task.
In contrast, satellite networks offer ubiquitous, seamless connectivity, coverage enhancement, and data offloading. Therefore, ISTNs leverage the strengths of both systems, ensuring high-quality, seamless connections across various environments \cite{SurveyTut_Roadto6G, ProIEEE_6GVision_Challenge_Opp,Zaid_PIMRC23}.
Additionally, due to the rapid increase in data demand and the proliferation of devices, 3GPP has highlighted the potential of ISTNs to operate TN and NTN systems in the same band, such as the C-band \cite{3gpp.38.863}. 
Due to long propagation and broadband properties, the C-band has been widely used for satellite services. Recently, these qualities made it highly attractive in 5G for good coverage and high data throughput, facilitated by the US FCC and RSPG in Europe \cite{FCC20_Cband_5G, RSPG23_5Gband}, and standardized by 3GPP for 5G NR.
The convergence of ISTN development and the global interest in the C-band envisions many benefits for future ISTNs. 
However, operating in the same band introduces critical cross-network interference issues in various environments. Urban environments, in particular, present complex characteristics that significantly impact the channel gain between satellites and ground segments \cite{Hung_MeditCom24}. Additionally, the movement of UEs and LEOSats introduces dynamics and time-varying changes \cite{VuHa_ICC23,Tewel_TCom24,Juan_PIMRC23,VuHa_PIMRC24}; hence, environmental characteristics and movements should be considered for ISTN designs.

The spectral coexistence of TN and satellite networks has been studied extensively \cite{Yeongi_Access20_Coexistence_mmWave, Eva_Access20, Sormunen22_ASMSSPSC, Okati24_Coexistence_Sband, 3gpp.38.863}. The authors in \cite{Yeongi_Access20_Coexistence_mmWave} investigated coexistence systems at the millimeter-wave frequency, analyzing co-channel and adjacent channel interference in NTNs. In \cite{Eva_Access20}, the authors analyzed the interference and out-of-band emission caused by 5G networks on fix-satellite-service (FSS) earth stations in C-band coexistence scenarios. The study in \cite{Sormunen22_ASMSSPSC} examined scenarios where TNs and NTNs operate in adjacent 5G NR bands in the S-band, analyzing the impact of adjacent channel interference on system performance. In \cite{Okati24_Coexistence_Sband}, the authors explored coexistence scenarios in the S-band, analyzing co-channel interference, network coverage, and achievable rates in different cases. Notably, the 3GPP recently discussed various coexistence scenarios between satellite and TNs in \cite{3gpp.38.863}, analyzing throughput loss in S-band. However, these analyses rely on statistical channels without capturing the geographical characteristics of complex environments.

Therefore, novel solutions considering urban-environment characteristics and terminal mobility to address critical cross-network interference are essential. This paper studies the ISTNs in an urban area operating in the C-band. Considering the mobility of UEs and LEOSats, we formulate a sum-rate (SR) maximization problem to optimize link selection for UEs, under the BS and LEOSat load constraints. This problem is challenging to solve directly due to the non-convex rate function. To address this efficiently, we employ the successive-convex-approximation (SCA) approach \cite{Hung_Access22, Hung_ICCE22, LinChen_TWC23, Hung_Access22UAV} to develop an iterative algorithm.
In the simulation, the ray-tracing method and an actual 3D map are utilized to capture urban characteristics. Additionally, the routes of the UEs are obtained using Google Navigator. The simulated heatmap of the link budget provides valuable insights into the effects of urban characteristics on ISTNs. For the max-SR problem, the numerical results demonstrate the efficiency of the proposed algorithm in terms of SR, compared to a proposed greedy-based algorithm.

\section{System Model}
\begin{figure}
    \centering
    \includegraphics[height=30mm, width=60mm]{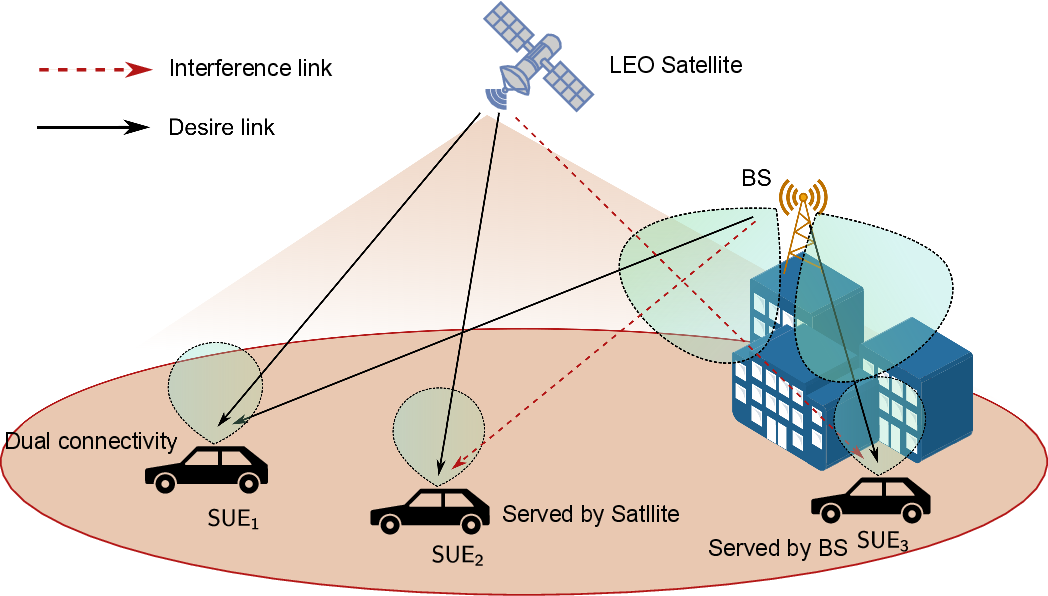}
    \captionsetup{font=small}
    \vspace{-1mm}
    \caption{System model of ISTNs.}
    \label{fig:SystemModel}
    \vspace{-2mm}
\end{figure}

\vspace{-1mm}
This work studies the downlink ISTN systems in the urban environments as depicted in Fig.~\ref{fig:SystemModel}, wherein LEOSats and TNs operate in the same frequency band, i.e., C-band, during a time-windows of $N_{TS}$ TSs. The system includes $M$ LEOSats, $N$ BSs, and $K$ UEs (cars).
We further assume that the UEs can be served by the LEOSat, one BS, or both at any time. 

\vspace{-1mm}
\subsection{Tx-UE Channel Model}
\vspace{-1mm}
Let $h_{n,k}^{t}$ and $g_{m,k}^{t}$ be the channel gain of ${\sf{BS}}_{n}-{\sf{UE}}_{k}$ and ${\sf{LEOSat}}_{m}-{\sf{UE}}_{k}$ links at TS $t$, respectively. The channel gain is calculated by the inverse of path-loss (PL).
By utilizing the multi-path channel model, the equivalent PL of Tx-UE (Tx is LEOSat/BS) link can be modeled as
\vspace{-2mm}
\beq \label{eq: PL model}
    L_{\sf{UE}}^{\sf{Tx}} \text{[dB]} = -\log_{10} \Big( \Big\vert \scaleobj{0.8}{\sum_{i=0}^{N_{\sf{ray}}}  \frac{G_{\sf{r}}(\theta_{i}^{\sf{a}},\varphi_{i}^{\sf{a}}) G_{\sf{t}}(\theta_{i}^{\sf{d}},\varphi_{i}^{\sf{d}})}{L_{i} L_{\sf{B}}} e^{-j \phi_{i}}} \Big\vert^{-1} \Big),  \vspace{-2mm}
\eeq
wherein $N_{\sf{ray}}$ is the number of propagation rays from the Tx to the UE. $L_{i}$ and $\phi_{i}$ are the propagation loss and phase delay of ray $i$. 
$G_{\sf{r}}(\cdot)$ and $G_{\sf{t}}(\cdot)$ are the antenna gain patterns of the UE and the Tx.
$(\theta_{i}^{\sf{a}},\varphi_{i}^{\sf{a}})$ the elevation and azimuth angles of arrive at the UE, and $(\theta_{i}^{\sf{d}},\varphi_{i}^{\sf{d}})$ are angles of departure at the Tx corresponding to ray $i$. 
Especially, ray $i, \; i \geq 1$, is the LoS ray or the ray reaching the UE by reflection or diffraction on the obstacles in the environment. Ray $i=0$ is the ray propagating through the wall to reach the UE, the loss component $L_{0}$ of this ray can be modeled $L_{0} = L_{\sf{FS}} + 2 L_{\sf{wall}}$, wherein $L_{\sf{FS}}$ is the free space path loss and $L_{\sf{wall}}$ is the penetration loss which is given in \cite{3gpp.38.901}.
In addition, $L_{\sf{B}}$ is the other basic loss defined by \cite{3gpp.38.901} and \cite{3gpp.38.811} for TN and NTN links, respectively.

Subsequently, the antenna gain pattern of LEOSats, BSs, and UEs are modeled as follows. Regarding the LEOSats, we employ the antenna gain pattern modeled by the Bessel function for that satellite given in \cite{3gpp.38.811}. The BS's antenna gain is modeled by the non-active antenna system pattern expressed in \cite{3gpp.38.863}. Additionally, the UE (the car) is assumed to be equipped with a patch antenna array on the rooftop, wherein its antenna pattern can be obtained by the MATLAB antenna toolbox.

\subsection{User Association Model}
\vspace{-1mm}
Regarding the BS-UE link, we introduce a binary variable $\boldsymbol{\alpha} \triangleq [\alpha_{n,k}^{t}]_{\forall (n,k,t)}$ to indicate the BS-UE association: $\alpha_{n, k}^{t} = 1 \text{ if } {\sf{UE}}_{k} \text{ connects to } {\sf{BS}}_{n} \text{ at TS } t$ and $ \alpha_{n, k}^{t}= 0 \text{ otherwise}$.
In addition, one assumes that each UE can connect to at most one BS at each TS, which is ensured by the constraint:
\vspace{-1mm}
\beq
    (C1): \quad \scaleobj{0.8}{\sum_{\forall n}} \alpha_{n, k}^{t} \leq 1, \forall (k,t).
    \vspace{-1mm}
\eeq
Additionally, one assumes that ${\sf{BS}}_{n}$ can serve at most $\psi_{n}^{\sf{B}}$ users. Let $\eta_{n}^{{\sf{B}},t}$ be the number of other connected users of ${\sf{BS}}_{n}$ at TS $t$, the number of connections at each BS can be ensured as
\vspace{-1mm}
\beq
    (C2): \quad \scaleobj{0.8}{\sum_{\forall k}} \alpha_{n, k}^{t} \leq \psi_{n}^{\sf{B}} - \eta_{n}^{{\sf{B}},t}, \forall (n,t),
    \vspace{-1mm}
\eeq
Subsequently, regarding the LEOSat-UE association, we introduce another binary variable $\boldsymbol{\beta} \triangleq [\beta_{m,k}^{t}]_{\forall (m,k,t)}$ such that $\beta_{m, k}^{t} =1 \text{ if } {\sf{UE}}_{k} \text{ connects to } {\sf{LEOSat}}_{m} \text{ at TS } t$ and $\beta_{m, k}^{t} =0 \text{ otherwise}$.
In addition, we assume that each UE can connect to at most one LEOSat at each TS, which is ensured as
\vspace{-1mm}
\beq
    (C3): \quad \scaleobj{0.8}{\sum_{\forall m}} \beta_{m, k}^{t} \leq 1, \forall (k,t).
    \vspace{-1mm}
\eeq
Assuming that ${\sf{LEOSat}}_{m}$ can serve at most $\psi_{m}^{\sf{S}}$ users at each TS, let $\eta_{m}^{{\sf{S}},t}$ be the number of other connected users to ${\sf{LEOSat}}_{m}$ at TS $t$, the UE connection number at each LEOSat obeys 
\vspace{-1mm}
\beq
    (C4): \quad \scaleobj{0.8}{\sum_{\forall k}} \beta_{m, k}^{t}  \leq \psi_{m}^{\sf{S}} - \eta_{m}^{{\sf{S}},t}, \forall (m,t).
    \vspace{-1mm}
\eeq
Furthermore, we assume that dual connectivity is supported, i.e., UE can connect to both BS and LEOSat simultaneously. To ensure seamless connectivity, one assumes that each UE has at least one connectivity at each TS, guaranteed by
\vspace{-1mm}
\beq \label{eq: QoE const}
    (C5): \quad \scaleobj{0.8}{\sum_{\forall n}} \alpha_{n, k}^{t} + \scaleobj{0.8}{\sum_{\forall m}} \beta_{m, k}^{t} \geq 1, \forall (k,t).
\eeq
\vspace{-5mm}
\subsection{Transmission Model}
We assume that components within the same networks are coordinated for interference mitigation.
Hence, one assumes that the intra-network interference is negligible and only cross interference between TNs and NTNs is considered. 
Consequently, if ${\sf{UE}}_{k}$ connects to ${\sf{BS}}_{n}$, its SINR  can be given as
\vspace{-1mm}
\beq \label{eq: SINR_n,k}
    \gamma_{n,k}^{{\sf{B}}, t}(\boldsymbol{\alpha}, \boldsymbol{\beta}) = \scaleobj{0.8}{\frac{\alpha_{n,k}^{t} P_{n}^{{\sf{B}}} h_{n,k}^{t}}{ \sum_{\forall m} (\eta_{m}^{\sf{S}, t} + \sum_{\forall k'} \beta_{m,k'}^{t} ) P_{m}^{{\sf{S}}} g_{m,k}^{t} + \sigma_{k}^2}}, \vspace{-1mm}
\eeq
wherein $\sigma_{k}^2$ is the AWGN power at ${\sf{UE}}_{k}$. $P_{n}^{\sf{B}}$ and $P_{m}^{\sf{S}}$ are the transmit power of ${\sf{BS}}_{n}$ and ${\sf{LEOSat}}_{m}$ for each of their connected users, respectively. These transmit powers are uniformly allocated as $P_{n}^{\sf{B}} = P_{n}^{\sf{B},max} / \psi_{n}^{\sf{B}}$ and $P_{m}^{\sf{S}} = P_{m}^{\sf{S},max} / \psi_{m}^{\sf{S}}$.
The throughput from ${\sf{BS}}_{n}$ to ${\sf{UE}}_{k}$ at TS $t$ can be computed as
\vspace{-1mm}
\beq \label{eq: Rate_n,k}
    R_{n,k}^{{\sf{B}}, t}(\boldsymbol{\alpha}, \boldsymbol{\beta}) = \log (1 + \gamma_{n,k}^{{\sf{B}}, t}(\boldsymbol{\alpha}, \boldsymbol{\beta}) ). \vspace{-1mm}
\eeq
Regarding the LEOSat-UE connection, if ${\sf{UE}}_{k}$ connects to ${\sf{LEOSat}}_{m}$, the corresponding SINR can be expressed as
\vspace{-1mm}
\beq \label{eq: SINR_m,k}
    \gamma_{m,k}^{{\sf{S}}, t}(\boldsymbol{\alpha}, \boldsymbol{\beta}) = \scaleobj{0.8}{\frac{\beta_{m,k}^{t} P_{m}^{{\sf{S}}} g_{m,k}^{t}}{ \sum_{\forall n} (\eta_{n}^{\sf{B}, t} + \sum_{\forall k'} \alpha_{n,k'}^{t} ) P_{n}^{{\sf{B}}} h_{n,k}^{t} + \sigma_{k}^2}}. \vspace{-1mm}
\eeq
The throughput of ${\sf{LEOSat}}_{m}\!-\!{\sf{UE}}_{k}$ link at TS $t$ is given as
\vspace{-1mm}
\beq \label{eq: Rate_m,k}
    R_{m,k}^{{\sf{S}}, t}(\boldsymbol{\alpha}, \boldsymbol{\beta}) = \log (1 + \gamma_{m,k}^{{\sf{S}}, t}(\boldsymbol{\alpha}, \boldsymbol{\beta}) ). \vspace{-1mm}
\eeq
Hence, the throughput of ${\sf{UE}}_{k}$ at TS $t$ can be expressed as
\vspace{-1mm}
\beq
    R_{k}^{t}(\boldsymbol{\alpha}, \boldsymbol{\beta}) = \scaleobj{0.8}{\sum_{n}} R_{n,k}^{{\sf{B}}, t}(\boldsymbol{\alpha}, \boldsymbol{\beta}) + \scaleobj{0.8}{\sum_{m}} R_{m,k}^{{\sf{S}}, t}(\boldsymbol{\alpha}, \boldsymbol{\beta}).
\eeq
\vspace{-6mm}
\subsection{Problem Formulation}
The UEs mobility, LEOSats movement, and complexity of urban areas lead to the time-varying change in the system. Therefore, the UEs need to change the connections between BSs/LEOSats over time to maximize their SR. 
The max-SR problem can be mathematically formulated as
\vspace{-1mm}
\bieq {ll} \label{eq: max SR}
    \max_{\boldsymbol{\alpha}, \boldsymbol{\beta}} \quad & \scaleobj{1.1}{\sum_{k, t}} R_{k}^{t}(\boldsymbol{\alpha}, \boldsymbol{\beta}) \quad \st \; (C1)-(C5).  \vspace{-1mm}
\eieq
It can be seen that problem \eqref{eq: max SR} is non-convex due to the non-convexity of the rate formulas in the objective. 

\vspace{-2mm}
\section{Proposed Sum-rate Maximization Solution}

Problem \eqref{eq: max SR} is difficult to solve directly due to the binary variable $\boldsymbol{\alpha}$  and $\boldsymbol{\beta}$, and the non-convexity of the objective function. Hence, we first relax the binary variables into the continuous ones as $\alpha_{n,k}^{t} \in [0,1], \forall (n,k,t)$ and $\beta_{m,k}^{t} \in [0,1], \forall (m,k,t)$, respectively. Besides, to convexify the rate functions in the objective function, let us perform the rate functions by their lower bounds $\lambda_{n,k}^{\sf{B},t}$ and $\lambda_{m,k}^{\sf{S},t}$ as
\vspace{-2mm}
\bieq{ll} \label{eq: lower bound rate}
    \lambda_{n,k}^{\sf{B},t} \leq R_{n,k}^{{\sf{B}}, t}(\boldsymbol{\alpha}, \boldsymbol{\beta}), \quad \forall n,k,t, \subnum \\
    \lambda_{m,k}^{\sf{S},t} \leq R_{m,k}^{{\sf{S}}, t}(\boldsymbol{\alpha}, \boldsymbol{\beta}), \quad \forall m,k,t. \subnum 
\eieq
However, constraint \eqref{eq: lower bound rate} is still non-convex. Subsequently, \eqref{eq: lower bound rate} can be convexified by the following proposition.
\begin{mypro} \label{pro: convexify rate}
   \textit{ Constraints in \eqref{eq: lower bound rate} can be convexified as}
   \vspace{-2mm}
    \bieq{ll} \label{eq: convexify rate}
        \log(\alpha_{n,k}^{t} P_{n}^{{\sf{B}}} h_{n,k}^{t} \!\! + \!\! \scaleobj{0.8}{\sum_{\forall m}} (\eta_{m}^{\sf{S}, t} \!\! + \!\! \scaleobj{0.8}{\sum_{\forall k'}} \beta_{m,k'}^{t} ) P_{m}^{{\sf{S}}} g_{m,k}^{t} \!\! + \!\! \sigma_{k}^2)  \geq \lambda_{n,k}^{\sf{B},t} \!\! + \!\! \mu_{k}^{\sf{B},t}, \non \\
        \scaleobj{1.02}{\sum_{\forall m}} (\eta_{m}^{\sf{S}, t} \!\! + \!\! \scaleobj{0.8}{\sum_{\forall k'}} \beta_{m,k'}^{t} ) P_{m}^{{\sf{S}}} g_{m,k}^{t} \!\! + \!\! \sigma_{k}^2 \leq e^{\mu_{k}^{\sf{B},t,(i)}}(\mu_{k}^{\sf{B},t}\!\! - \!\!\mu_{k}^{\sf{B},t,(i)} \!\!+ \!\!1), \non \\
        \log(\beta_{m,k}^{t} P_{m}^{{\sf{S}}} g_{m,k}^{t} \!\!+\!\! \scaleobj{0.8}{\sum_{\forall n}} (\eta_{n}^{\sf{B}, t} \!\! + \!\! \scaleobj{0.8}{\sum_{\forall k'}} \alpha_{n,k'}^{t} ) P_{n}^{{\sf{B}}} h_{n,k}^{t} \!\!+\!\! \sigma_{k}^2) \!\!\geq\!\! \lambda_{m,k}^{\sf{S},t} \!\!+\!\! \mu_{k}^{\sf{S},t}, \non \\
        \scaleobj{1.02}{\sum_{\forall n}} (\eta_{n}^{\sf{B}, t} \!\! + \!\! \scaleobj{0.8}{\sum_{\forall k'}} \alpha_{n,k'}^{t} ) P_{n}^{{\sf{B}}} h_{n,k}^{t} \!\!+\!\! \sigma_{k}^2 \leq e^{\mu_{k}^{\sf{S},t,(i)}}(\mu_{k}^{\sf{S},t} \!\! - \!\! \mu_{k}^{\sf{S},t,(i)} \!\! + \!\! 1). \quad \vspace{-3mm}
    \eieq
\end{mypro}
\begin{IEEEproof}
        Consider function $f(x,y) = \log(1+\frac{x}{y + c})$, $x,y \geq 0, c > 0$, constraint $f(x,y) \geq u$ is rewritten as
    \begin{subnumcases}{}
        \log(x + y + c) \geq u + z, \label{eq: apx rate 1a} \\
        \log(y + c) \leq z, \label{eq: apx rate 1b}
    \end{subnumcases}
    where $z$ is a slack variable. Taking the exponent operator to both sides of \eqref{eq: apx rate 1b} 
    and applying first-order approximation to the exponential term yield the following convex constraints
    \begin{subnumcases}{}
        \log(x + y + c) \geq u + z, \label{eq: apx rate 3a} \\
        y + c \leq e^{z^{(i)}}(z - z^{(i)} + 1), \label{eq: apx rate 3b}
    \end{subnumcases}
    wherein $z^{(i)}$ is a feasible $z$ value at iteration $i$.
    By setting $ x \!=\!\! \alpha_{n,k}^{t}P_{n}^{{\sf{B}}} h_{n,k}^{t}, \;
    y \!=\! \sum_{\forall (m,k')} \beta_{m,k'}^{t} P_{m}^{{\sf{S}}} g_{m,k}^{t},$ $
    c = \sum_{\forall m}\eta_{m}^{\sf{S}, t}P_{m}^{{\sf{S}}} g_{m,k}^{t} + \sigma_{k}^2,
    $ and $u = \lambda^{\sf{B},t}_{n,k}$, we obtain two first convex constraints in \eqref{eq: convexify rate}. 
    Subsequently, by setting 
    $
    x = \beta_{m,k}^{t} P_{m}^{{\sf{S}}} g_{m,k}^{t}, \;
    y = \sum_{\forall (n,k')} \alpha_{n,k'}^{t} P_{n}^{{\sf{B}}} h_{n,k}^{t}, \;
    c = \sum_{\forall n} \eta_{n}^{\sf{B}, t}  P_{n}^{{\sf{B}}} h_{n,k}^{t} + \sigma_{k}^2,
    $ and $u = \lambda^{\sf{S},t}_{m,k}$, we obtain two last convex constraints in \eqref{eq: convexify rate}.
\end{IEEEproof}
Thanks to Proposition~\ref{pro: convexify rate}, we obtain the successive convex problem at iteration $i$ around feasible point $(\boldsymbol{\mu}^{{\sf{B}}, {(i)}}, \boldsymbol{\mu}^{{\sf{S}}, {(i)}})$ as
\bieq {ll} \label{eq: convex prob}
    \max_{\substack{\boldsymbol{\alpha}, \boldsymbol{\beta}, \boldsymbol{\lambda}^{\sf{B}},  \\ \boldsymbol{\lambda}^{\sf{S}}, \boldsymbol{\mu}^{\sf{B}}, \boldsymbol{\mu}^{\sf{S}}}} \; \scaleobj{0.8}{\sum_{k, t}}  (\scaleobj{0.8}{\sum_{\forall n}} \lambda_{n,k}^{\sf{B},t} + \scaleobj{0.8}{\sum_{\forall m}} \lambda_{m,k}^{\sf{S},t} ) \;\;
     \st \; (C1)\!\!-\!\!(C6), \eqref{eq: convexify rate}. \quad 
     \vspace{-2mm}
\eieq
By iterative solving problem \eqref{eq: convex prob}, we can obtain the solution for \eqref{eq: max SR}. However, the obtained $\boldsymbol{\alpha}$ and $\boldsymbol{\beta}$ by solving \eqref{eq: convex prob} are continuous. Their binary solution can be recovered as \cite{VuHa_TVT16}
     \vspace{-2mm}
\bieq{ll} \label{eq: recover binary}
       \hspace{-10mm} \text{if } \alpha_{n,k}^{t} \! \geq \! 1/2, \alpha_{n,k}^{t} \!=\! 1, \text{ else }  \alpha_{n,k}^{t} \! = \!0,  \subnum \\
       \hspace{-10mm} \text{if } \beta_{m,k}^{t} \! \geq \! 1/2, \beta_{m,k}^{t} \!=\! 1, \text{ else }  \beta_{m,k}^{t} \! = \!0,  \subnum \vspace{-3mm}
\eieq
The proposed algorithm is summarized in Algorithm~\ref{alg_1}. 

\begin{algorithm}[!t]
\footnotesize
	\begin{algorithmic}[1]
 \captionsetup{font=small}
		\protect\caption{\textsc{Proposed Iterative Algorithm}}
		\label{alg_1}
        \STATE Set $i=0$ and generate an initial point $(\boldsymbol{\mu}^{{\sf{B}}, {(0)}}, \boldsymbol{\mu}^{{\sf{S}}, {(0)}})$.\\
		\REPEAT
		\STATE Solve \eqref{eq: convex prob} to obtain $(\boldsymbol{\alpha}^{\star}, \boldsymbol{\beta}^{\star}, \boldsymbol{\lambda}^{\sf{B},{\star}},  \boldsymbol{\lambda}^{\sf{S},{\star}}, \boldsymbol{\mu}^{\sf{B},{\star}}, \boldsymbol{\mu}^{\sf{S},{\star}})$.
		\STATE Update $ (\boldsymbol{\mu}^{{\sf{B}}, {(i)}}, \boldsymbol{\mu}^{{\sf{S}}, {(i)}}) := (\boldsymbol{\mu}^{{\sf{B}}, \star}, \boldsymbol{\mu}^{{\sf{S}}, \star})$ and set $i=i+1$.
		\UNTIL Convergence
        \STATE Recovery variables $\boldsymbol{\alpha}$ and $\boldsymbol{\beta}$ by \eqref{eq: recover binary}.
        \STATE \textbf{Output:} The solution $(\boldsymbol{\alpha}^{\star}, \boldsymbol{\beta}^{\star})$.
    \end{algorithmic}
    \normalsize 
\end{algorithm}


\section{Simulation and Implementation} \label{sec:CINR results}
\vspace{-1mm}
This section conducts the simulation of the considered ISTNs in an urban area. First, we carry out the simulation to obtain the channel gain of UEs over time and the link budget in terms of carrier-to-interference-plus-noise-ratio (CINR) for the entire examined area. Especially, to reflect the complexity of the urban area, the ray-tracing method and actual 3D map are employed to compute the multi-path components in \eqref{eq: PL model}.
Subsequently, based on the obtained channel gain, the numerical results for the SR problem are provided. 
The simulation is conducted in an area in London City.
The map is divided into multiple map segments with size of $(0.005^\circ \times 0.005^\circ)$ in latitude and longitude, respectively, and the receiver grid with size of $(100 \times 100)$ in each map segment is considered for CINR assessment in the entire map. Additionally, one BS is deployed in each map segment at the highest building. The UE routes are obtained from the Google Navigator application. The key simulation parameters are summarized in Table.~\ref{tab: parameter}.

\setlength{\textfloatsep}{3pt}
\begin{figure}
\begin{subfigure}{0.5\textwidth}
    \centering
    \includegraphics[width=70mm]{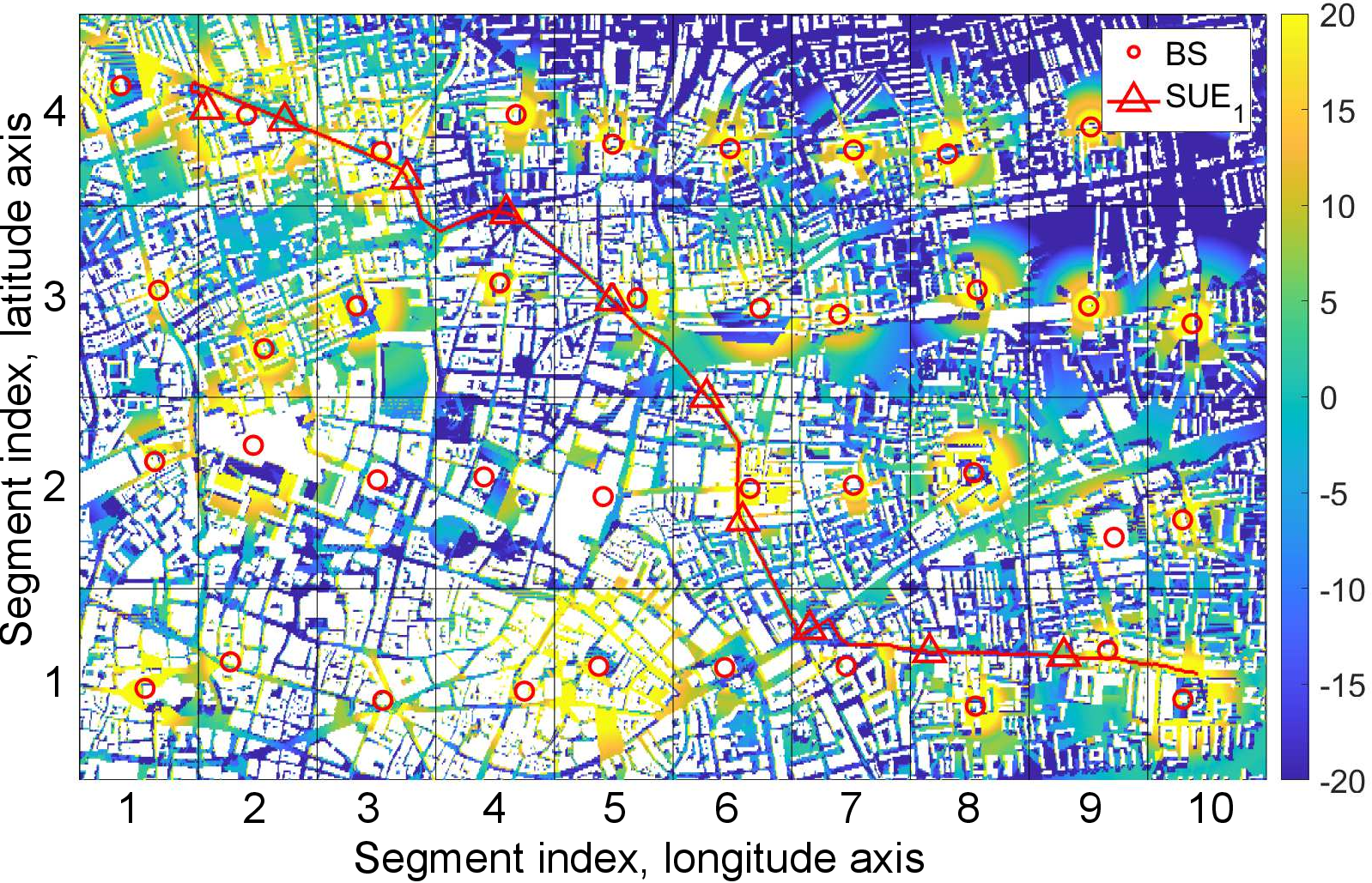}
    \captionsetup{font=small}
    \vspace{-1.2mm}
    \caption{Snapshot CINR heatmap of BS-UE link and UE trajectory.}
    \label{fig:CINR_BSSUE_heatmap_SUE1}
\end{subfigure}
\begin{subfigure}{0.5\textwidth}
    \centering
    \includegraphics[width=70mm]{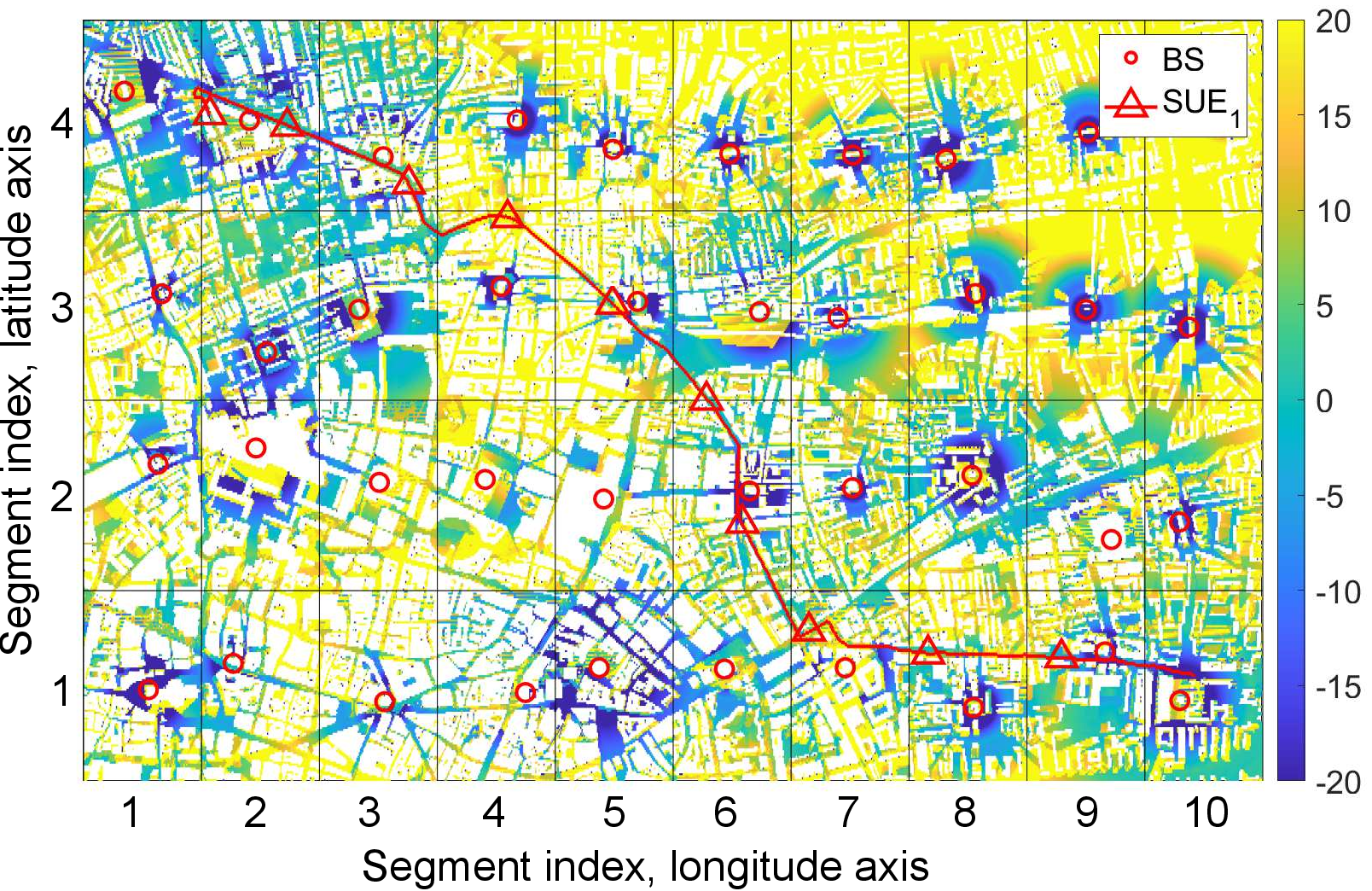}
    \vspace{-1.2mm}
    \caption{Snapshot CINR heatmap of LEOSat-UE link and UE trajectory.}
    \captionsetup{font=small}
    \label{fig:CINR_SatSUE_heatmap_SUE1}
\end{subfigure}
\captionsetup{font=small}
\vspace{-5mm}
\caption{Heatmap of CINR in dB scale with a UE trajectory.}
     \label{fig:CINR_SUE_trajec}
\end{figure}

\begin{figure*}[t]
    \centering
    \includegraphics[width=170mm]{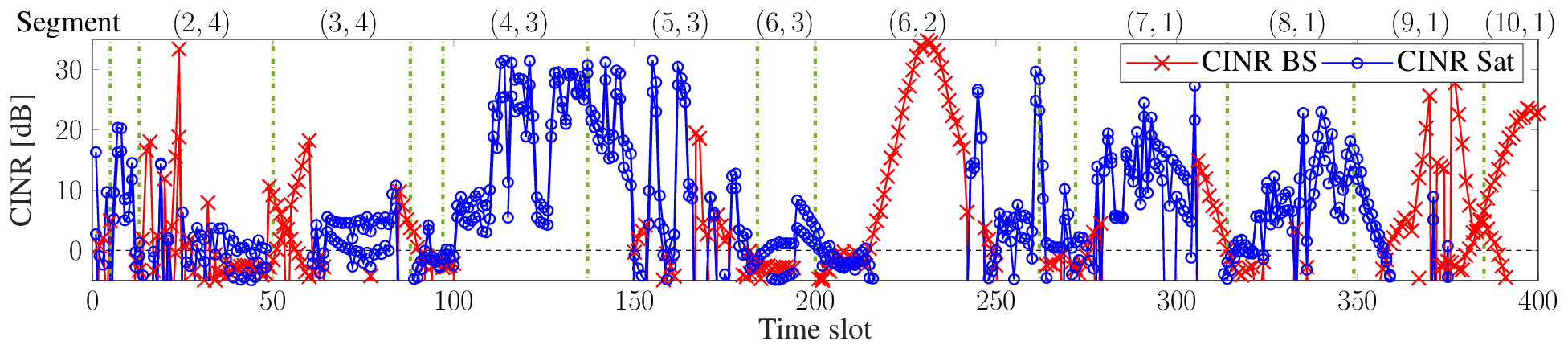}
    \vspace{-2mm}
    \captionsetup{font=small}
    \caption{CINR of links from LEOSats and BSs to a UE over TSs.}
    \label{fig:CINR_time_SUE1}
    \vspace{-4mm}
\end{figure*}

\begin{table}[!t]
\captionsetup{font=footnotesize}
	\caption{Simulation Parameters}
	\label{tab: parameter}
	\centering
		\scalebox{0.8}{
	\begin{tabular}{l|l}
		\hline
		Parameter & Value \\
		\hline\hline
        Operation frequency, $f_c$              & $3.4$ GHz \\
        System bandwidth, $B$   & $20$ MHz \\
        Latitude limitation of area             & $[51.5115 ^\circ \text{N}, 51.5315^\circ \text{N}]$   \\
        Longitude limitation of area            & $[0.1022^\circ \text{W} , 0.0522^\circ \text{W}]$   \\
        Size of one map segment                   & $0.005^\circ \times 0.005^\circ$ \\
        LEO altitude                  & $500$ km \\
        LEO inclination angle   & $53^\circ$ \\
        Number of LEOSats at each time  & $2$ \\
        LEOSat antenna parameters    & \cite{3gpp.38.811} \\
        Maximum transmit power of LEOSat, $P^{\sf{S,max}}_{m}$    &   $16$ dBW \\
        Receiver point grid size in one segment & $100 \times 100$ points      \\
        Maximum transmit power of BS, $P^{\sf{B,max}}_{n}$    &   $42$ dBm \\
        Number of UEs, $K$ &   $4$ \\
        UE patch antenna array size         & $2 \times 2$\\
        Maximum UE antenna gain         & $12$ dBi \\
        UE receiver noise figure, $G_f$ & $1.2$ dB \\
        UE antenna temperature, $T_a$  & $150$ K \\
        Maximum number served users at BSs, $\psi_{n}^{\sf{B}}$ & $20$ \\
        Maximum number served users at LEOSats, $\psi_{m}^{\sf{S}}$ & $80$ \\
		\hline		   				
	\end{tabular}}
 \vspace{-2mm}
\end{table}

     \vspace{-1mm}
\subsection{Simulation Results for Link Budget Assessment} \label{sec:CINR simul results}
\vspace{-1mm}
Based on the outcome of the ray-tracing simulation, we can calculate the equivalent PL in \eqref{eq: PL model}. Based on the PL, the CINR is computed simply as in \cite{Guidotti_ICC20_NTNlink}. Subsequently, the CINR heatmaps of BS-UE and LEOSat-UE links at a TS across the entire map are shown as in  Fig.~\ref{fig:CINR_BSSUE_heatmap_SUE1} and \ref{fig:CINR_SatSUE_heatmap_SUE1}, respectively. In addition, the trajectory of a UE is also depicted in these figures. 

Regarding the BS-UE CINR in Fig.~\ref{fig:CINR_BSSUE_heatmap_SUE1}, the CINR varies across the locations on the map. In particular, the BS-UE CINR at locations far from BSs and with low building density is very low, about from $-10$ dB to $-20$ dB, while that at locations near to BSs and with high building density is very high, exceeding $15$ dB. This phenomenon is due to the impact of UE antenna directivity and blocking signals by buildings. Particularly, the increase in distance between BS-UE leads to the lower elevation at UE toward BSs, hence, the antenna gain decreases as per the patch antenna pattern. Additionally, at the location with lower building density, the LEOSat-UE link is less blocked, which results in a decrease in the BS-UE CINR.
Besides, the LEOSat-UE CINR has the opposite trend with that of BS-UE CINR as in Fig.~\ref{fig:CINR_SatSUE_heatmap_SUE1}. The LEOSat-UE CINR increases as BS-UE distance increases and the building is more sparse. This trend is also due to the influence of the UE antenna pattern and the blockage of buildings.

In addition, the UE usually moves across different segments on the map, as the depicted example UE route in Fig.~\ref{fig:CINR_SUE_trajec}. As a result, the CINR values change corresponding to the UE location. For an insight into the time-varying CINR as UE moving, Fig.~\ref{fig:CINR_time_SUE1} shows the BS-UE and LEOSat-UE CINR metrics over time corresponding to the UE route in Fig.~\ref{fig:CINR_SUE_trajec}. Compare between the CINR overtime in Fig.~\ref{fig:CINR_time_SUE1} and the CINR heatmaps in Fig.~\ref{fig:CINR_SUE_trajec}, it can be seen the significant changes in CINR due to the influence of the environment, especially the buildings across the locations on the map.


\vspace{-2mm}
\subsection{Numerical Results for Max-SR Problem}
\vspace{-1.2mm}
For comparison purposes, we further proposed a simple \textbf{greedy algorithm} as in Algorithm~\ref{alg: greedy}. For convenience, the MATLAB notation is used for the index of matrices.


\begin{algorithm}[t]
\footnotesize
\begin{algorithmic}[1]
     \captionsetup{font=small}
    \protect\caption{\textsc{Greedy Algorithm}}
    \label{alg: greedy}
    \STATE \textbf{Input:} Parameters $\{ \psi_{n}^{\sf{B}}, \psi_{m}^{\sf{S}}, \eta_{n}^{{\sf{B}},t}, \eta_{m}^{{\sf{S}},t} \}_{\forall (n,m,t)}$ and channel gain matrices $ \mathbf{h}=[h_{m,k}^{t}]_{\forall (m,k,t)} $ and $ \mathbf{g} = [g_{m,k}^{t}]_{\forall (n,k,t)}$
    \STATE \textbf{Initialize:} Zero matrices $\boldsymbol{\alpha}$ and $\boldsymbol{\beta}$
    \FOR{Each TS $t$}
    \REPEAT
    \STATE Find the index of the maximum element in $\mathbf{h}(:,:,t)$: $(\hat{n},\hat{k})$
    \STATE \textbf{if} {$\scaleobj{0.8}{\sum_{\forall k}} \alpha_{\hat{n}, k}^{t}  \leq \psi_{\hat{n}}^{\sf{B}} - \eta_{\hat{n}}^{{\sf{B}},t}$}
        \textbf{then} Update $\alpha_{\hat{n},\hat{k}}^{t} = 1$
         and $\mathbf{h}(:,\hat{k},t) = \mathbf{0}$
        \STATE \textbf{else} Update $\mathbf{h}(\hat{n},:,t) := \mathbf{0}$
    \UNTIL All BSs or UEs are assigned 
    \REPEAT
    \STATE Find the index of the maximum element in $\mathbf{g}(:,:,t)$: $(\hat{m},\hat{k})$
    \STATE \textbf{if} {$\scaleobj{0.8}{\sum_{\forall k}} \beta_{\hat{m}, k}^{t}  \leq \psi_{\hat{m}}^{\sf{S}} \!- \eta_{\hat{m}}^{{\sf{S}},t}$}
        \textbf{then} Update $\beta_{\hat{m},\hat{k}}^{t} = 1$
         and $\mathbf{g}(:,\hat{k},t) = \mathbf{0}$
        \STATE \textbf{else} Update $\mathbf{g}(\hat{m},:,t) := \mathbf{0}$
    \UNTIL All LEOSats or UEs are assigned 
    \ENDFOR
    \STATE \textbf{Output:} Link selection solution $\boldsymbol{\alpha}, \boldsymbol{\beta}$
\end{algorithmic} 
\normalsize
\end{algorithm}

\begin{figure}[!t]
    \centering
    \includegraphics[width=80mm]{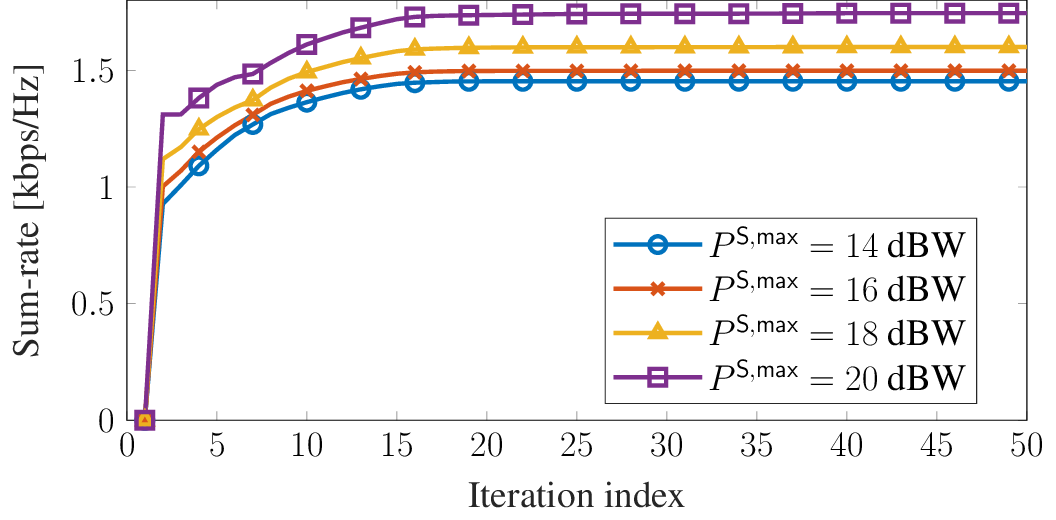}
    \vspace{-2mm}
    \captionsetup{font=small}
    \caption{The convergence rate of sum-rate over iterations.}
    \label{fig:SR_convergence_conf}
    \vspace{-3mm}
\end{figure}

Fig.~\ref{fig:SR_convergence_conf} depicts the SR convergence of Algorithm 1 in different cases of maximum LEOSat transmit power. It can be seen that the SR increases quickly and reaches the saturation value after only about $20$ iterations. In particular, at $P^{\sf{S},max} = 14, 16, 18$ and $20$ dBW, Algorithm 1 requires about $16, 16, 17$ and $19$ iterations for convergence, respectively.

\begin{figure}[!t]
    \centering
    \includegraphics[width=80mm]{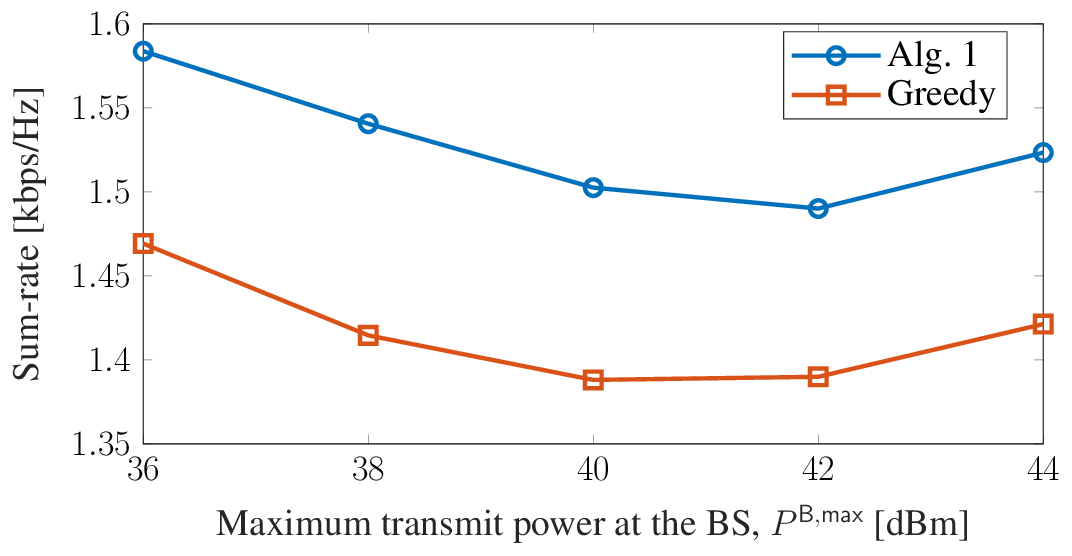}
    \vspace{-2mm}
    \captionsetup{font=small}
    \caption{Sum-rate versus the maximum transmit power at the BS.}
    \label{fig:SR_PBS_conf}
    \vspace{-1mm}
\end{figure}

Fig.~\ref{fig:SR_PBS_conf} depicts the SR versus the maximum transmit power at BSs of two algorithms. Interestingly, the SR provided by both algorithms has the same trend, it decreases and then increases as $P^{\sf{B,max}}$ increases. This phenomenon can be explained as follows. Increasing the transmit power at the BS causes cross-system interference to the signal sent from LEOSats to UEs, leading to the degradation in SR. However, exceeding a certain transmit power value, the throughput offered by BSs dominates that of the LEOSat-UE links, resulting in an increase in SR. Compared between the two algorithms, in all considered values of $P^{\sf{B,max}}$, Algorithm 1 always provides better SR performance than that of the greedy algorithm. The performance gap between two lines is significant, about $0.11$ and $0.1$ kbps/Hz at $P^{\sf{B,max}} = 36$ dBm and $P^{\sf{B,max}} = 42$ dBm, respectively.

\begin{figure}
    \centering
    \includegraphics[width=80mm]{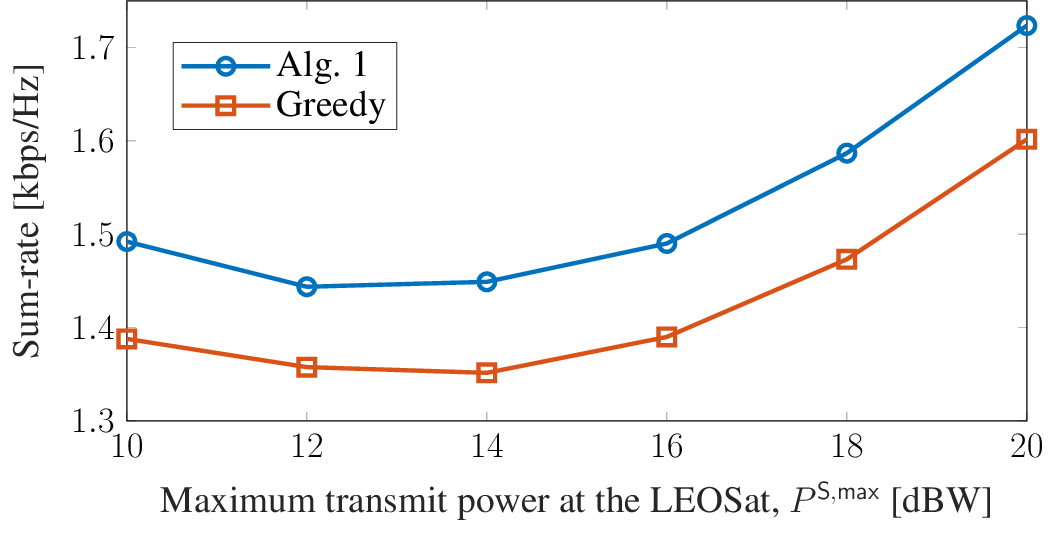}
    \captionsetup{font=small}
    \vspace{-2mm}
    \caption{Sum-rate versus the maximum transmit power at the LEOSat.}
    \label{fig:SR_PSat_conf}
   \vspace{0mm}
\end{figure}

Fig.~\ref{fig:SR_PSat_conf} illustrates the SR performance at different LEOSat power budgets. The SR evolution trend and the explanation are similar to those in Fig.~\ref{fig:SR_PBS_conf}. In addition, at $P^{\sf{S,max}}$ larger than $14$ dBW, the throughput offered by LEOSats seems to dominate that of BS-UE links clearly since the SR increases quickly as $P^{\sf{S,max}}$ increases in this range. In particular, Algorithm 1 provides the SR of about $1.45$ kbps/Hz and $1.73$ kbps/Hz at $P^{\sf{S,max}} = 14$ dBW and $P^{\sf{S,max}} = 20$ dBW, respectively. Furthermore, Algorithm 1 outperforms the greedy algorithm in all considered cases, the gap between the two lines is about $0.1$ kbps/Hz. These numerical results have demonstrated the efficiency of Algorithm 1 compared to the greedy algorithm.

\section{Conclusion}
This work studied the time-window ISTNs in realistic urban environments wherein cross-system interference is considered. The system is highly dynamic due to the UE and LEOSat movements. Hence, We formulated an SR maximization problem to optimize UE link selection over time. To address this efficiently, we employed the SCA method to develop an iterative algorithm. In the simulation, to capture the effect of urban area characteristics on the channels from LEOSats and BSs to UEs, the ray-tracing method and realistic 3D map are utilized. The resulting CINR heatmap offered valuable insights into the impact of urban environments on ISTNs. Numerical results demonstrated that our proposed algorithm outperformed the greedy algorithm in terms of SR performance.

 \vspace{-1mm}
 \section*{Acknowledgment}
 \vspace{-1mm}
 \small
 This work has been supported by the Luxembourg National Research Fund (FNR) under the project INSTRUCT (IPBG19/14016225/INSTRUCT) and project MegaLEO (C20/IS/14767486).
 \normalsize

\vspace{-1mm}
\bibliographystyle{IEEEtran}
\bibliography{Journal}
\end{document}